# Band in ARPES caused by photodissociation of Landau-Pekar polarons

A.E.Myasnikova, E.N. Myasnikov

South Federal University, 344090 Rostov-on-Don, Russia

We consider decay of phonon condensate into phonons at photodissociation of the strong coupling large polaron (SCLP), when the charge carrier becomes free. Expression to calculate the band in ARPES caused by photodissociation of SCLP is obtained. The band in ARPES caused by photodissociation of strong-coupling large-radius polarons is a broad band with the shape determined by Poisson distribution. It can be structured or unstructured depending on the phonon dispersion since a distance between neighbouring lines comprising the band is the phonon energy. Half-width of the band is in the interval $1.3 - 1.7 E_p$, depending on the phonon energy. The band maximum is situated approximately at the electron energy $E_{phot}-W-3.2E_p$ (where $E_{phot}$ is the photon energy, W is work function), and its position does not depend on the electron wave vector direction.

## 1. Introduction

We have shown earlier [1] that the strong-coupling large radius polaron (i.e. polarons that occur at strong electron-phonon interaction in a medium where the conduction band is not narrow) contains phonon condensate. (In other words, the polarization field in such polarons is in quantum-coherent state.) If the charge carrier is removed from the SCLP on its photodissociation the phonon condensate decays into phonons. As any Bose-condensate, the phonon condensate is a superposition of states with different number of quanta where summands are phased up to small quantum fluctuations. Therefore photodissociation of such polarons is accompanied by radiation of different number of phonons in each act. Decay of the phonon condensate at the SCLP photodissociation results in a wide band in the optical conductivity spectrum. The calculated band in the optical conductivity spectrum [2] has the maximum at about $4.2E_p$ and half-width about $2.5 E_p$, where $E_p$ is the polaron binding energy. The calculated band is in good conformity with mid-IR band in the optical conductivity spectra of complex oxides. The predicted ratio (≈4) of the maximum position of the band caused by the SCLP photoionization to the frequency of maximum of the band caused by phototransitions into polaron excited states calculated by other group [3] is also in good conformity with experiments.

Obviously, decay of phonon condensate at the SCLP photodissociation should display itself as sufficiently wide band in photoemission too. The aim of the present study is to obtain an expression to calculate the band in ARPES spectrum caused by photodissociation of SCLP.

## 2. Photodissociation of the strong-coupling large polaron

Following Emin [] we calculate the band in the optical conductivity spectrum caused by photodissociation of SCLP using the simplest treatment of the photoelectric effect. Therefore our result can be easily reconstructed in order to describe the band in ARPES caused by the SCLP photodissociation. So, the initial state of the system is a charge carrier in autolocalized state (coupled in the polarization potential well that represents a polarization field in coherent state.

The polaron photodissociation occurs as a result of interaction of an electromagnetic wave of frequency $\Omega$ with the charge carrier in the polaron (the longitudinal field of the polarization in the polaron obviously does not interact with the transverse electromagnetic wave). The operator of the interaction has the form

$$\hat{H}_{int} = \frac{e\hbar(\mathbf{kA})}{m^*c} e^{i\mathbf{Q}\cdot\mathbf{r}}, \qquad (1)$$

where $\hbar\mathbf{k}$ is the electron momentum, $\mathbf{A}$ is the amplitude of the vector potential of the electromagnetic field, related with its intensity $I$ as it follows: $I = \Omega \mathbf{A}^2/2\pi\hbar c$; $\mathbf{Q}$ is the wave vector of the electromagnetic wave. According to Fermi golden rule a probability of transition of the system from the state $|i\rangle$ into the state $|f\rangle$ per unit time under the influence of the operator $\hat{H}_{int}$ has the form

$$W_{if} = \frac{2\pi}{\hbar}\left|\langle f|\hat{H}_{int}|i\rangle\right|^2 \delta(E_i - E_f), \qquad (2)$$

where $E_i$ and $E_f$ are the energies of the initial and final states of the whole system. If the initial state is the ground state [1,2] of the polaron in an electromagnetic field of a frequency $\Omega$ then

$$|i\rangle = \sqrt{\beta^3/7\pi}(1+\beta r)\exp(-\beta r)\prod_{\mathbf{q}}|d_{\mathbf{q}}\rangle \qquad (3)$$

and $E_i = -E_p - W + \hbar\Omega$ where $E_p$ is the polaron binding energy, W is work function, and Pekar wave function for the carrier state in the polaron is used.

To describe the final state we following Emin [3] use the simplest treatment of the photoelectric effect where the final state of the charge carrier is approximated as a free-carrier state. The vectors of possible final states of the phonon field are the eigen vectors $|\{v_{\mathbf{q}}\}\rangle = \prod_{\mathbf{q}}|v_{\mathbf{q}}\rangle$ of the non-shifted Hamiltonian $\hat{H}_{ph} = \sum_{\mathbf{q}}\hbar\omega_{\mathbf{q}} b_{\mathbf{q}}^+ b_{\mathbf{q}}$ describing the states with the certain number of quanta $v_{\mathbf{q}}$ in each harmonics. Thus, after the photodissociation the state (3) transforms into a state

$$|f\rangle = L^{-3/2}\exp(i\mathbf{kr})\prod_{\mathbf{q}}|\{v_{\mathbf{q}}\}\rangle, \qquad (4)$$

provided the sum of $v_{\mathbf{q}}$ (taking values 0 or 1) from the set $\{v_{\mathbf{q}}\}$ yields a certain number $v$. Hence, the energy of the final state is $E_f = \frac{\hbar^2\mathbf{k}^2}{2m^*} + \hbar\omega v$, if we neglect the dependence of $\omega$ on $\mathbf{q}$. Thus,

$$\delta(E_i - E_f) = \delta\left(-E_p + \hbar\Omega - W - \frac{\hbar^2\mathbf{k}^2}{2m^*} - v\hbar\omega\right). \qquad (5)$$

As the operator $\hat{H}_{int}$ acts only on the electron variables, the matrix element of the transition has the form

$$\langle f|\hat{H}_{int}|i\rangle = \int d\mathbf{r} L^{-3/2}\exp(-i\mathbf{kr})\hat{H}_{int}\sqrt{\beta^3/7\pi}(1+\beta r)e^{-\beta r}\prod_{\mathbf{q}}\langle v_{\mathbf{q}}|d_{\mathbf{q}}\rangle. \qquad (6)$$

Naturally, it contains the scalar product of the vector of a coherent state of the phonon field by a vector of its state with the certain number of phonons. After carrying out the integration in (18) the probability of the electron transition into a state with the wave vector with modulus $k$ and direction in a spatial angle $\sin\theta\,d\theta\,d\varphi$ has the form

$$dW_{\{v_{\mathbf{q}}\},\mathbf{k}} = \frac{2\pi}{\hbar}\left\{\frac{e\hbar(\mathbf{kA})}{m^*c}32\sqrt{\frac{\pi}{7\beta^3}}L^{-3/2}\left(1+\beta^{-2}|\mathbf{Q}-\mathbf{k}|^2\right)^{-3}\right\}^2 \cdot \prod_{\mathbf{q}}|\langle v_{\mathbf{q}}|d_{\mathbf{q}}\rangle|^2 d\rho(\mathbf{k}), \quad (7)$$

where

$$d\rho(\mathbf{k}) = \frac{m^*L^3 k(\varepsilon)}{(2\pi)^3 \hbar^2}\sin\theta\, d\theta\, d\varphi = \frac{m^*L^3 k(\varepsilon)}{(2\pi)^3 \hbar^2}d\Omega \quad (8)$$

is the spectral density of the final carrier states with the wave vector directed in the body angle dΩ [3]. According to (5) the electron momentum $\hbar\mathbf{k}$ and energy $\varepsilon$ in the final state are related as follows:

$$\hbar k(\varepsilon) = \sqrt{2m^*\varepsilon} = \sqrt{2m^*(\hbar\Omega - E_p - \nu\hbar\omega)}. \quad (9)$$

According to the energy and the momentum conservation laws (Exp.(5) and $\mathbf{Q} = \mathbf{k} + \mathbf{q}_0$, where $\mathbf{q}_0$ is the wave vector of the phonon field after the polaron photoionization, $\mathbf{q}_0 = \sum_{\mathbf{q}}\mathbf{q}v_{\mathbf{q}}$) an experiment can measure only the probability (7) summarized over all possible sets $\{v_{\mathbf{q}}\}$ having the same values of $\nu = \sum_{\mathbf{q}}v_{\mathbf{q}}$ and $\mathbf{q}_0$:

$$dW_{\{v_{\mathbf{q}}\},\mathbf{k}} = \frac{2\pi}{\hbar}\left\{\frac{e\hbar(\mathbf{kA})}{m^*c}32\sqrt{\frac{\pi}{7\beta^3}}L^{-3/2}\left(1+\beta^{-2}|\mathbf{Q}-\mathbf{k}|^2\right)^{-3}\right\}^2 \cdot \sum_{\{v_{\mathbf{q}}\}}^{\nu}\prod_{\mathbf{q}}|\langle v_{\mathbf{q}}|d_{\mathbf{q}}\rangle|^2 d\rho(\mathbf{k}), \quad (10)$$

Here symbol $\nu$ over $\Sigma$ denotes that the sum is carried out over the sets $\{v_{\mathbf{q}}\}$ satisfying the condition $\sum_{\mathbf{q}} v_{\mathbf{q}} = \nu$. Besides, there is not a set with $\nu = 0$ among the sets $\{v_{\mathbf{q}}\}$. Indeed, in such a case $\mathbf{q}_0 = 0$, hence, $\mathbf{k} = \mathbf{Q}$, and $\mathbf{kA} = \mathbf{QA} = 0$, i.e. the probability of a transition with appearance of such a set $\{v_{\mathbf{q}}\}$ is zero.

Summarized probability (10) contains as a multiplier the sum

$$P_\nu = \sum_{\{v_{\mathbf{q}}\}}^{\nu}\prod_{\mathbf{q}}|\langle v_{\mathbf{q}}|d_{\mathbf{q}}\rangle|^2. \quad (11)$$

The sum (11) calculated in [2] with taking into account the fact that $d_k \ll 1$ has the form:

$$P_\nu = \sum_{\{v_{\mathbf{q}}\}}^{\nu}\prod_{\mathbf{q}}|\langle v_{\mathbf{q}}|d_{\mathbf{q}}\rangle|^2 = \frac{\overline{\nu}^{\nu-1}}{(\nu-1)!}e^{-\overline{\nu}} \quad (12)$$

where $\overline{\nu}$ is the average number of phonons radiated at the SCLP photodissociation [2]:

$$\overline{\nu} = 2E_p(\hbar\omega)^{-1}. \quad (13)$$

To write the scalar product in (10) let us use ordinary for ARPES experiment geometry [4] where the wave vector $\mathbf{Q}$ of the incident photon lies in XZ plane of the coordinate system and makes angle ψ with z axes. The wave vector $\mathbf{k}$ of the electron is considered to have a component $k_{||}$ lying in the XY plane of the coordinate system (and coinciding with the sample surface) and $k_\perp$ in the perpendicular direction. Then

$$(\mathbf{kA}) = k_{||}\cos\varphi A\cos\psi + \sqrt{2m^*\varepsilon/\hbar^2 - k_{||}^2}\, A\sin\psi \quad (14)$$

where the relation $k_\perp = \sqrt{2m^*\varepsilon/\hbar^2 - k_{||}^2}$ is taken into account. Finally, the wave vector $\mathbf{Q}$ of the photon ordinarily used in ARPES experiments can be approximated as zero [].

Thus, expression (10) takes the form

$$dW_{\{v_q\},\mathbf{k}} = \frac{256 e^2}{7\pi\hbar m^* c^2 \beta^3} \frac{\left(k_{||}\cos\varphi\cos\psi + \sqrt{2m^*\varepsilon/\hbar^2 - k_{||}^2}\sin\psi\right)^2 A^2 k(\varepsilon)}{(1+\beta^{-2}k^2)^6} \cdot \frac{\bar{v}^{v-1}}{(v-1)!} e^{-\bar{v}} d\Omega$$

(15)

Exp.(15) represents the probability of the polaron photodisociation with appearance of ν phonons and electron with the energy ε and wave vector **k** in the body angle dΩ around certain direction. The direction is determined by **k** projection on X-, Y- and Z-axes $k_{||}\cos\varphi$, $k_{||}\sin\varphi$, and $k_\perp = \sqrt{2m^*\varepsilon/\hbar^2 - k_{||}^2}$, respectively. In Exp.(15) k(ε) is determined by Exp.(9).

Figs.1,2 demonstrate calculated by Exp.15 band in ARPES caused by photodissociation of SCLP with the binding energy 0.14 eV and 0.17 eV, respectively. The upper curves on both figures correspond to $k_x$=1, $k_y$=0, the lower curves correspond to $k_x$=0, $k_y$=0. For both figures the electron-phonon interaction constant α=6 (it determines the phonon energy for given $E_p$), $\hbar\Omega - W$ =20eV and phonon dispersion is neglected. If do not neglect the phonon dispersion the points on Figs.1,2 will transform into "partial" bands, and the resulting summarized band can be structured or unstructured depending on the phonon dispersion since a distance between neighbouring lines comprising the band is the phonon energy. Half-width of the band is in the interval 1.3 - 1.7$E_p$, depending on the phonon energy. The band maximum is situated approximately at the electron energy $E_{phot}$-W-3.2$E_p$ (where $E_{phot}$ is the photon energy, W is work function), and its position does not depend on the electron wave vector direction.

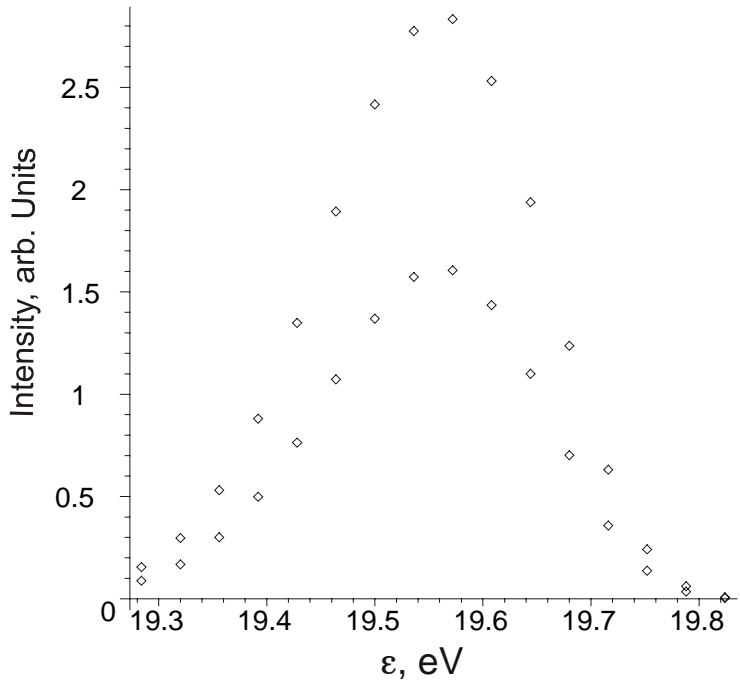

Fig.1. Band in ARPES caused by photoionization of SCLP with $E_p$=0.14 eV in neglect of phonon dispersion, at α=6 and hΩ-W=20 eV. The upper and lower "curves" correspond to kx=1 and kx=0, respectively, ky=0.

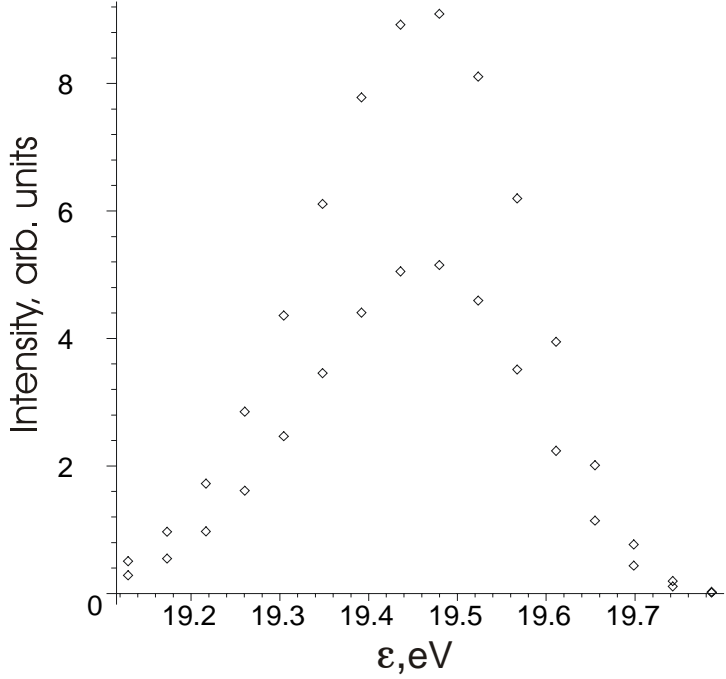

Fig.2. Band in ARPES caused by photoionization of SCLP with $E_p$=0.17 eV in neglect of phonon dispersion, at α=6 and hΩ-W=20 eV. The upper and lower "curves" correspond to kx=1 and kx=0, respectively, ky=0.

It is worth noting that intensity of the so-called "coherent" zero-phonon line (band) is equal to zero as it can be seen from Figs1,2. The reason of this is explained below Exp.(10).

For example, we can calculate the polaron binding energies in $YBa_2Cu_3O_{6+y}$, $Nd_2CuO_{4-y}$, $La_2CuO_{4+y}$ and $La_{2-x}Sr_xCuO_{4+y}$ from the position of maximum of the mid-infrared band in their optical conductivity spectra [5]. Then the maximum of the bands in ARPES of these substances caused by photodissociation of the polarons will be approximately at the electron energy (with respect to the energy $E_{phot}$-W) -0.48eV, -0.52eV, -0.44eV and -0.4eV, respectively.

Table 1

|  | $\hbar\Omega_{max}^{opt.cond.}$,eV [5] | $E_p$, eV | $\hbar\Omega_{max}^{ARPES}$, eV, with respect to $E_{phot}$-W |
|---|---|---|---|
| $Yba_2Cu_3O_{6+y}$ | 0.62±0.05 | ≈0.155 | ≈0.48 |
| $Nd_2CuO_{4-y}$ | 0.76±0.01 | ≈0.17 | ≈0.52 |
| $La_2CuO_{4+y}$ | 0.6±0.02 | ≈0.14 | ≈0.44 |
| $La_{2-x}Sr_xCuO_{4+y}$ | 0.53±0.05 | ≈0.13 | ≈0.4 |

To make comparison of the theoretical results with experiments Figs.3,4 demonstrate ARPES spectra of underdoped $La_{2-x}Sr_xCuO_4$ [6] and $Nd_{2-x}Ce_xCuO_4$ [7], respectively.

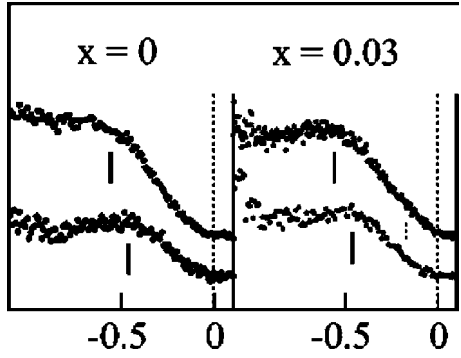
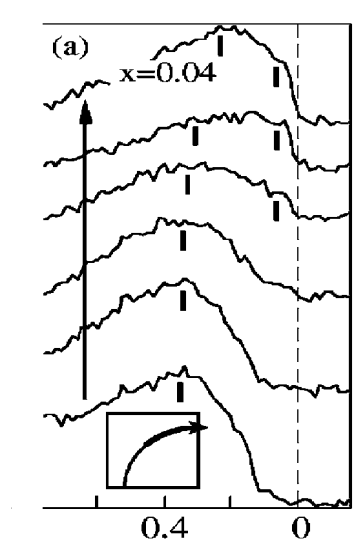

Fig.3. ARPES spectra of underdoped La$_{2-x}$Sr$_x$CuO$_4$ [6]

Fig.4. ARPES spectra of underdoped Nd$_{2-x}$Ce$_x$CuO$_4$ [7]

3. Temperature behavior of the band caused by SCLP photoionization

Velocity of SCLP is limited by the minimum phase (or maximum group) velocity u of phonons participating in SCLP formation. (It is shown for phonon dispersion of the form $\omega^2 = \omega_0^2 + u^2 k^2$ in [8]). As a result temperature of SCLP thermal destruction is much lower than their binding energy E$_p$ [9]. The temperature corresponding to double lowering of the polaron concentration due to their thermal destruction can be approximated as [10]

$$T_c = 0.278 E_p \left( \frac{m^* u c}{p_0} \right)^{0.176}$$

where m* is the bare carrier effective mass, u is the maximum group (and minimum phase) velocity of phonons interacting with the charge carrier, $c = 1/\varepsilon_0 - 1/\varepsilon_\infty$, $\varepsilon_0$ is a static dielectric constant, $\varepsilon_\infty$ is a dielectric permittivity at high frequencies, p$_0$ - «maximum» momentum of the carrier in the polaron [9].

An example of temperature dependence of SCLP concentration is demonstrated by Fig.5 [10].

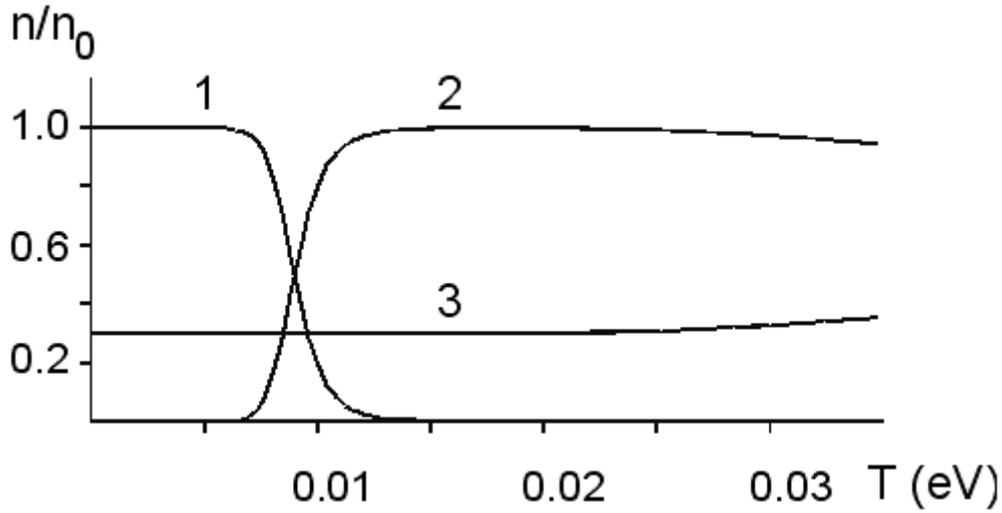

Fig.5. Temperature dependence of polaron concentration (curve 1), concentration of free carriers with the momentums $p<p_0$ (curve 2) and concentration of free carriers with the momentums $p>p_0$ (curve 3) in a system with total carrier concentration $n=1.3n_0$, polaron binding energy $E_p=0.11eV$, $u=5*10^4 cm/s$. $n_0$ is the maximum polaron concentration, $n_0 = 2*(4/3)\pi p_0^3/(2\pi\hbar)^3$.

Accordingly, integral intensity of the band caused by SCLP photodissociation will decrease as it occurs, e.g., in optical conductivity spectra of $\beta - Na_{0.33}V_2O_5$ shown by Fig.6 [11].

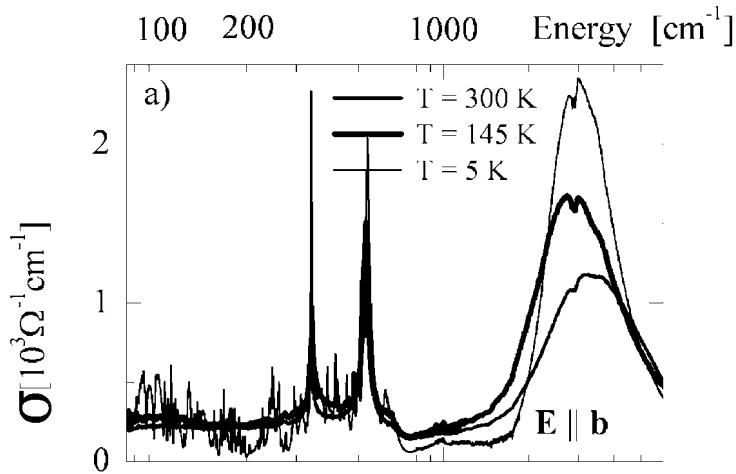

Fig.6. Optical conductivity spectra of $\beta - Na_{0.33}V_2O_5$ [11].

References

1. E. N. Myasnikov, A.E. Myasnikova, and F. V. Kusmartsev, Phys. Rev. B **72**, 224303 (2005).
2. E. N. Myasnikov, A.E. Myasnikova, Z.P. Mastropas, cond-mat/0703693.